\documentclass[preprint]{elsarticle}

\usepackage[utf8]{inputenc}
\usepackage[T2A]{fontenc}

\usepackage{graphicx}
\usepackage{amsmath}
\usepackage{amssymb}
\usepackage{color}
\usepackage{epstopdf}

\journal{Applied Mathematical Modelling}
\begin{document}

\begin{frontmatter}

\title{Modelling and characterization of a pneumatically actuated peristaltic micropump}

\author{T. N. Gerasimenko\corref{cor1}}
\address{M.V. Lomonosov Moscow State University, Faculty of Physics, Moscow, Russia}
\ead{gerasimenko@physics.msu.ru}
\cortext[cor1]{Corresponding author}

\author{O. V. Kindeeva, V. A. Petrov, A. I. Khaustov \corref{cor2}}
\address{Moscow Aviation Institute, Moscow, Russia}

\author{E. V. Trushkin \corref{cor2}}
\address{Hemule GmbH, Berlin, Germany}

\begin{abstract}
There is an emerging class of microfluidic bioreactors which possess long-term, closed circuit perfusion under sterile conditions with \textit{in vivo}-like flow parameters. Integrated into microfluidics, peristaltic-like pneumatically actuated displacement micropumps are able to meet these requirements. We present both a theoretical and experimental characterization of such pumps. In order to examine volume flow rate, we have developed a mathematical model describing membrane motion under external pressure. The viscoelasticity of the membrane and hydrodynamic resistance of the microfluidic channel have been taken into account. Unlike other models, the developed model includes only the physical parameters of the pump and allows the estimation of their impact on the resulting flow. The model has been validated experimentally.
\end{abstract}

\begin{keyword}
mathematical model\sep micropump\sep volume flow rate
\MSC[2010] 74D05\sep 74F10
\end{keyword}

\end{frontmatter}

%

%\linenumbers
\section{Introduction}
\label{intro}
Various microfluidic systems have been recently designed to recreate an \textit{in vivo} microenvironment for cell cultures with regard to physiological mechanical stimulation \cite{Huh2010b, Huh2011}.
A pump providing the circulation of liquid is the key component of these devices \cite{Wu2010} and can be either external or integrated into a microfluidic chip \cite{Laser2004, Iverson2008, Huang2013}. A pump with a pulsatile flow creates significant variations in volume flow rate, affecting the cells by time-varying pressure and shear stress on the cells surface \cite{Jacobs1998, Hsiai2002, Pauw2000, Huang2013, Lin2014}. Since the manufacturing of the pump is a complicated process, it is highly desired to understand how the parameters of the pump affects the resulting flow already at the stage of the design.
In general, fluid-structure interaction modelling can only be addressed by numerical simulations \cite{Chakraborty2012, Osman2015}. The lumped element method is traditionally used to simplify modelling when the pump is analyzed separately from the microfluidic system and its characteristics are taken as boundary conditions for a liquid flow simulation within a microchannel \cite{Goldschmidtboing2005, Yang2015a}. In those cases when the law of motion of actuators is known, the problem reduces to solving the Navier-Stokes equation in a domain with moving boundaries, which can be done both with the help of classical mesh methods of computational hydrodynamics and Stokeslets-meshfree computations \cite{Aboelkassem2012, Aboelkassem2015}.

A representation of the pump as an equivalent electric circuit is the most common approach for pump modelling \cite{Bourouina1996, Morganti2005, Goulpeau2005, Huang2006}. These methods result in flow parameters close to the experimental ones, although the relation between the pump's physical parameters and the equivalent capacitance, resistance and inductivity is not clear enough.
Thus the effect of pump characteristics on flow properties can only be roughly estimated. Moreover, specialized software is needed when the equivalent scheme is complex and branched \cite{Bourouina1996, Bendib2001, Morganti2005}. 

\begin{figure}[!ht]
\begin{center}
\includegraphics{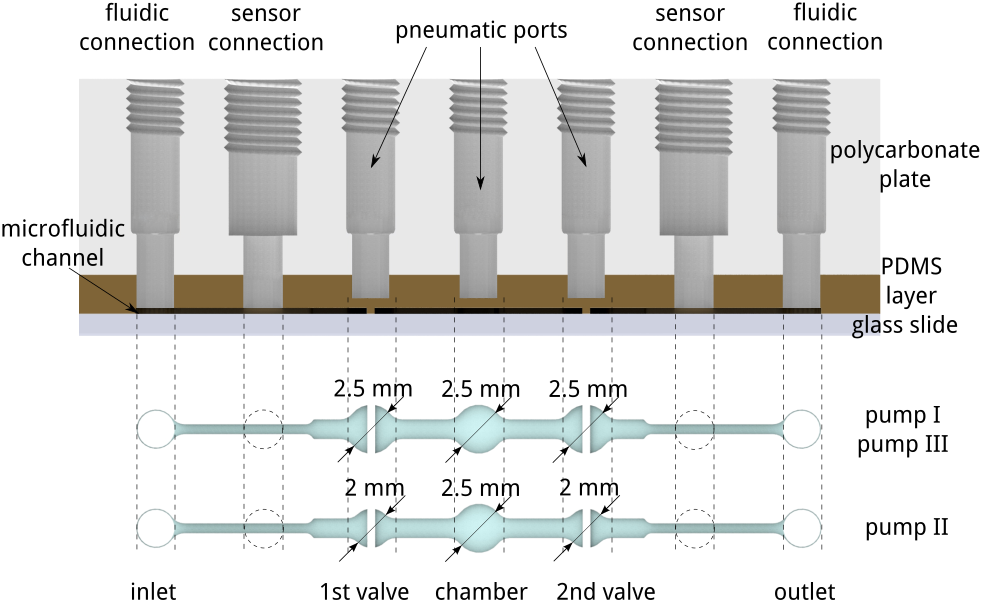}
\caption{The studied micropumps. All three pumps have a chamber diameter of 2.5 mm and a valve membrane thickness of 530$\pm$5 $\mu$m . Pump I and pump III have valve diameters of 2.5 mm. Pump II has valves with diameters of 2 mm. Pump I and pump II have chamber membrane thicknesses of 530$\pm$5 $\mu$m. Pump III has a chamber membrane with a thickness of 440$\pm$5 $\mu$m.}
\label{fig:pump}
\end{center} 
\end{figure}

\begin{figure}[!ht]
\begin{center}
\includegraphics{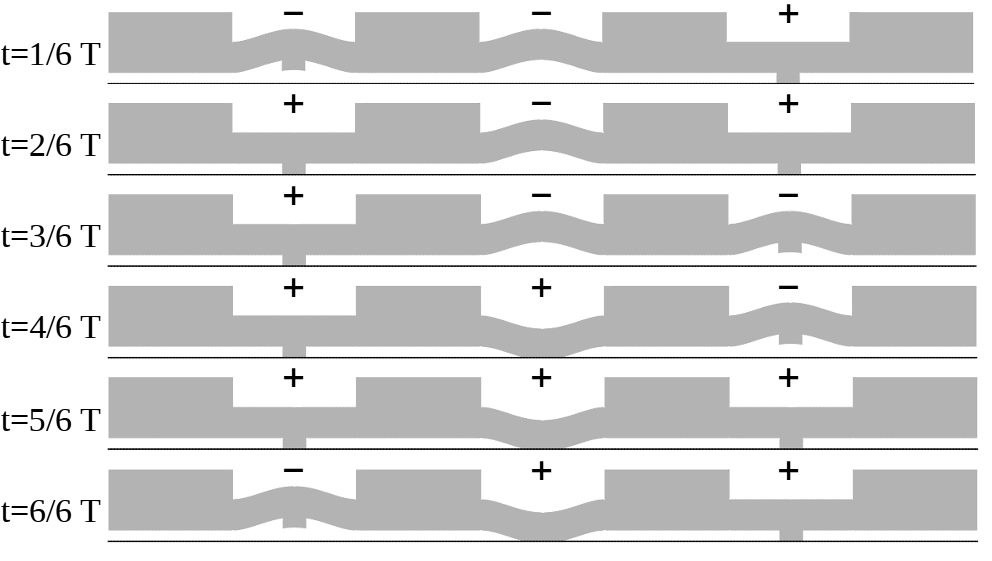}
\caption{The 6-step algorithm. The six steps together correspond to a single working cycle. The operating frequency is a reciprocal value of one step duration.}
\label{fig:motion}
\end{center}
\end{figure}

In our research we have studied a peristaltic-like displacement micropump consisting of two active valves with a working chamber in between. The valves and the working chamber are elastic, pneumatic-actuated membranes organized in a row along a microfluidic channel. Each valve membrane have a strap blocking the channel when the applied pressure is greater than atmospheric pressure (Fig.~\ref{fig:pump}). The valves and the chamber are operated by the proprietary Hemule control unit making the pressure over the membranes either higher (closed valve) or lower (opened valve) than one atmosphere according to the 6-step algorithm presented in Figure~\ref{fig:motion}. 

A theoretical model of variation of gas pressure in pneumatic chamber of a similar pump has been presented in \cite{Nedelcu2007}. The authors implicitly assume that the inflow of gas to the pneumatic chamber is isothermal and consider the relationship between the gas volume flow rate and the pressure drop to be linear; the corresponding proportionality coefficient is found numerically from a computational fluid dynamics calculation.
A model of an analogous pump has been proposed in \cite{Busek2013}, where the authors used the Hagen-Poiselle law to relate the air mass flow rate to the pressure drop. In that work, the motion of the pump membranes was also modelled, taking into account the membrane elasticity and fluid resistance within the channel. However, an expression for static deformations of the membrane was used, i.e., effectively, the deformation process was assumed to be quasi-static. This approach results in the unknown coefficients entering the final expression for the mean velocity of the fluid, whose values have to be determined empirically. Therefore, the model presented in \cite{Busek2013} cannot be used to analyze the pump at the design stage.

In our model we describe the variation of gas pressure over the pump membranes taking into account the presence of a throttle at the control unit outlet. The flowing of the gas is assumed to be adiabatic and is characterized by semiempirical relations which are well known in engineering science \cite{Kolovsky1999}.  When calculating the deflection of the membrane, like in \cite{Busek2013}, we take into account the resistance of the channel, but the displacement of the membrane from the quiescent position is described employing the dynamic equations of motion. The latter allows us to avoid the presence of opaque empirical coefficients in the resulting formulae. To compare the proposed model with experiment, we use data from pressure sensors and perform a simultaneous measurement of the average flow rate.

\section{Matherials and methods}
\begin{figure}
\begin{center}
\includegraphics{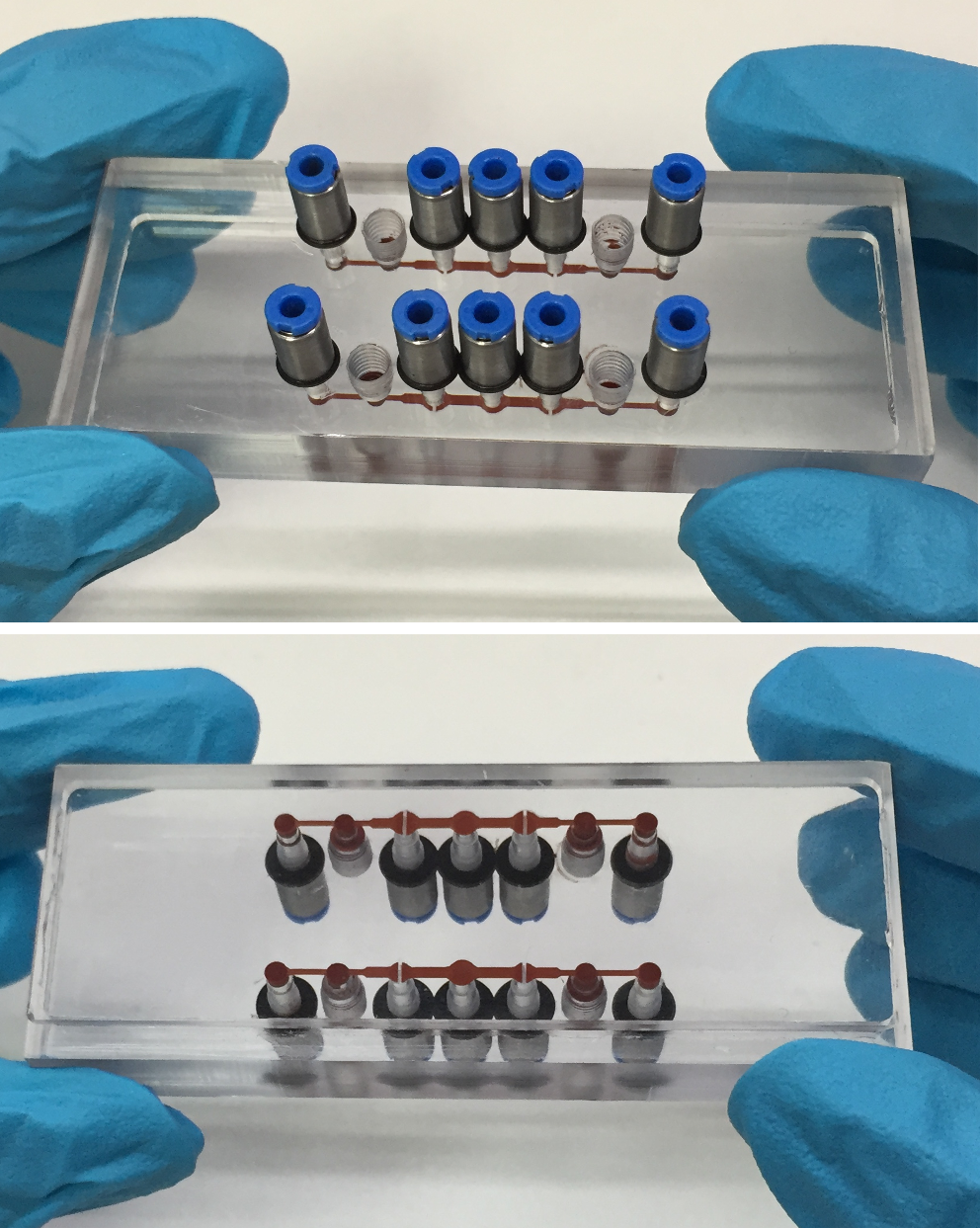}
\caption{Photo of the device with pumps I and II}
\label{fig:pump_photo}
\end{center}
\end{figure}

The micropumps are built in 2 mm polydimethylsiloxane (PDMS) layer using soft litography \cite{Unger2000} and then bonded to a standard 1-mm microscopic glass slide by means of plasma treatment \cite{Owen1994}. The other side of the PDMS layer is based on 10 mm polycarbonate plate with ports for fluidic and pneumatic connections as well as for pressure sensors (Fig.~\ref{fig:pump}). In our manufacturing process, we use PDMS with 10:1 monomer-to-curing agent ratio. Three versions of the pump have been made to analyze the impact of the pump geometry on its volume flow rate.
Two of them (pump I and pump III) have valve diameters of 2.5 mm. Pump II has smaller valves with diameters of 2 mm. All three pumps have a chamber diameter of 2.5 mm (Fig.~\ref{fig:pump_photo}). The thickness $h$ of the membranes is 530$\pm$5 $\mu$m except for the chamber membrane of pump III (440$\pm$5 $\mu$m). 
The pneumatic chamber diameters over the membranes are 2 mm and the channel height is 100 $\mu$m for all pumps.

\begin{figure}
\begin{center}
\includegraphics{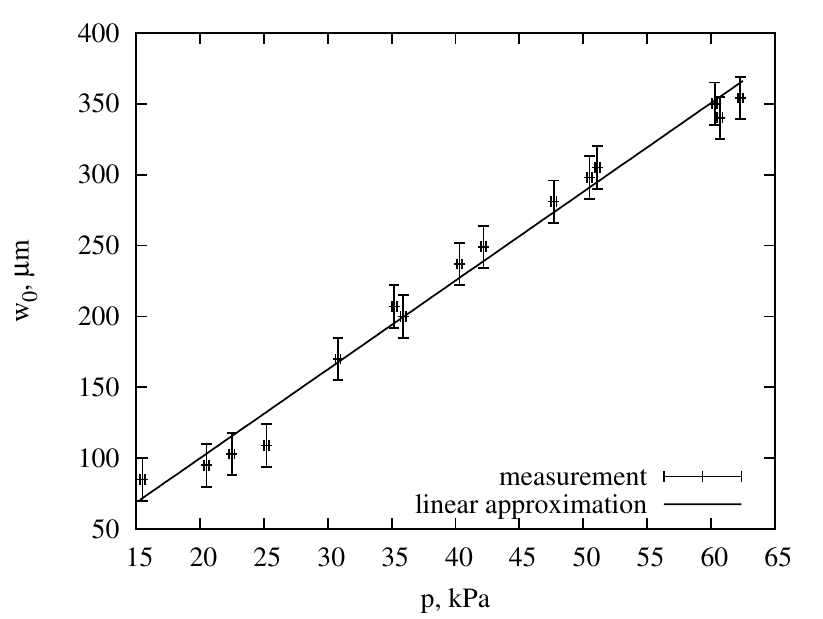}
\caption{The experimental dependency of the membrane center deflection on the pressure value}
\label{fig:Young}
\end{center}
\end{figure}

The stress-strain relationship of PDMS is highly dependent on its preparation. It is known that at intermediate temperatures it exhibits a behavior similar to a rubbery solid, and various nonlinear models have been proposed for its description \cite{Vanlandingham2005, Huang2005, Fincan2015, Gidde2017}. Mechanical and rheological properties of PDMS were investigated by Schneider et al. \cite{Schneider2009}, who showed that PDMS possesses a constant elastic modulus for strains up to 45\%. Huang and Anand \cite{Huang2005} demonstated a linear stress-stretch dependence of PDMS for stretches up to 1.3 for 5:1 monomer-to-curing agent ratio and up to 1.8 for 20:1 ratio. Based on these results, we infer that the linear model is valid between 1.3 and 1.8 for our 10:1 ratio. We assume PDMS has a constant Young modulus and adopt the Kelvin-Voigt model to describe the viscoelasticity similar to \cite{Fincan2015}. Thus the stress-strain relation is given by:

\begin{equation}
\sigma(t)=E\varepsilon(t)+\eta\dot{\varepsilon}(t)
\label{eq:KelvinVoigt}
\end{equation}
where $\sigma(t)$ is stress, $\varepsilon(t)$ is strain, $E$ is the PDMS Young modulus and $\eta$ is its viscosity. Since the Young modulus of PDMS depends on the method and conditions of the material preparation \cite{Johnston2014}, we estimate its value by measuring the deflection of the chamber membrane under a static load \cite{Reddy2006}:
\begin{equation}
E=\frac{3}{16}\frac{R^4}{h^3}(1-\nu^2)\frac{p}{w_0}
\end{equation}
where $R$ is the membrane radius, $\nu=0.499$ is the Poisson's ratio \cite{Johnston2014}, $w_0$ is the deflection of the membrane center from its quiescent position, and $p$ is the pressure over the membrane. In order to measure $w_0$, the pump is placed sideways on a stage of the inverted microscope. The static pressure over the working chamber membrane with a thickness of 440 $\mu$m and diameter of 2.5 mm is adjusted using the control unit. The obtained relationship between the membrane center deflection and the pressure value is approximated by a linear function (Fig.~\ref{fig:Young}).

The obtained Young modulus of 760$\pm$140 kPa is in good agreement with the results of \cite{Yang2015a} and \cite{Armani1999}.

\section{Mathematical model}
The proposed mathematical model is based on the idea that the volume flow rate caused by each membrane motion can be determined by integrating the membrane velocity over its surface due to the no-slip conditions on the membrane boundary. 
The velocity is obtained by solving a membrane's equation of motion. Since each membrane is actuated by air pressure in pneumatic tubes we first describe the time-dependency of the air pressure.

\subsection{Time-dependence of the applied pressure}
\label{sec:pressure}

Further for convenience we are using the gauge pressure instead of the absolute one. The positive and negative pressures inside the control unit are kept constant, therefore the transient processes of changing pressure are considered as an efflux of air from/to an infinite reservoir according to \cite{Kolovsky1999}:
\begin{equation}
\frac{d\zeta_{up}}{dt}= A \zeta_{up}^{2/7}\sqrt{\zeta_{up}^{-2/7}-1}
\end{equation}
\begin{equation}
\frac{d\zeta_{dn}}{dt}=-A \zeta_{dn}^{5/7}\sqrt{1-\zeta_{dn}^{2/7}}
\end{equation}
for increase and decrease of the pressure accordingly. Here, $\zeta_{up}= p_{up}/p_0$ and $\zeta_{dn}= p_{dn}/p_0$, where $p_0$ is the pressure provided on the control unit outlets, $p_{up/dn}$ describe the increasing and decreasing pressures correspondingly. The coefficient $A$ is calculated as follows:
\begin{equation}
A = \frac{R_z\sqrt{T_k}b\tilde{S}\tilde{\mu}}{V}
\end{equation}
$R_z=287$ J/(kg\,K) is the specific gas constant for dry air, $T_k=298$ K is the absolute temperature of the air, $b=0.155$~(kg K/J)$^{1/2}$ and $\tilde{\mu}=0.8$ are empirical coefficients \cite{Kolovsky1999}, $\tilde{S}=0.28$ mm$^{2}$ is the cross-section of the control unit's throttle orifice, $V=2261$ mm$^{3}$ is the total volume of the pneumatic tube and the connection.

The solution of these equations gives the dependence of increasing pressure on time in an implicit form:
\begin{multline}
(1-\zeta_{up}^{2/7})^{1/2}-\frac{2}{3}(1-\zeta_{up}^{2/7})^{3/2}+\frac{1}{5}(1-\zeta_{up}^{2/7})^{5/2}\\
-(1-\zeta_0^{2/7})^{1/2}+\frac{2}{3}(1-\zeta_0^{2/7})^{3/2}-\frac{1}{5}(1-\zeta_0^{2/7})^{5/2}
=-\frac{A}{7}(t-t_0)
\label{eq:p_up}
\end{multline}
and the decreasing one in an explicit form:
\begin{equation}
\zeta_{dn}=\left[ 1 - \left( \sqrt{1-\zeta_0^{2/7}} + \frac{A}{7}(t-t_0) \right)^2\right]^{7/2}
\label{eq:p_dn}
\end{equation}
with $\zeta_0$ being a normalized pressure value at $t_0$ for each case.

\begin{figure}
\begin{center}
\includegraphics{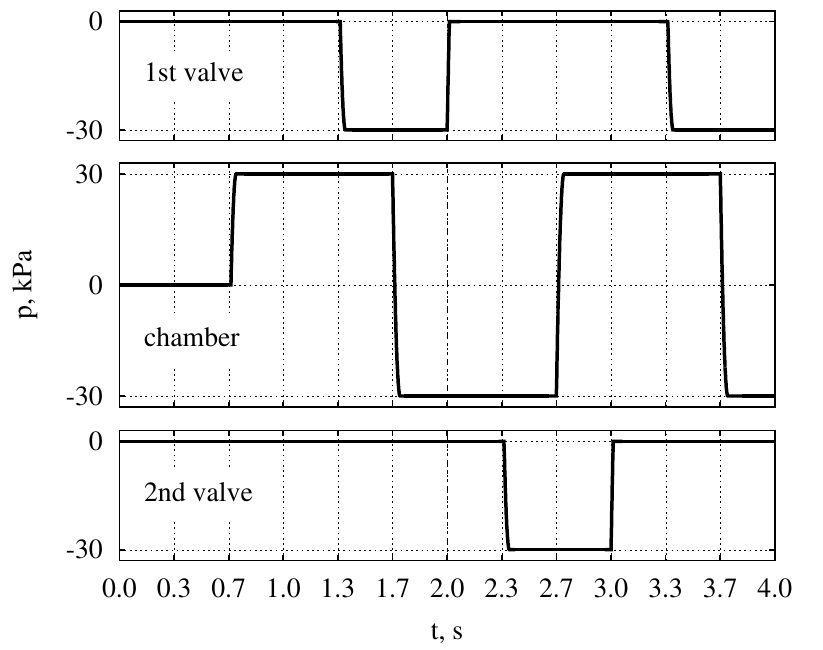}
\caption{The dependency of the pressure over the valves and chamber membranes on time for $p_0=30$~kPa and $f=3$~Hz control unit mode. The dashed line represents the beginning of the 6-step algorithm.}
\label{fig:pressures}
\end{center}
\end{figure}

The pneumatic signal edges $\Delta t_{up}$ and $\Delta t_{dn}$ are obtained setting $\zeta_{up}(t_0+\Delta t_{up})=1$ and $\zeta_{dn}(t_0+\Delta t_{dn})=0$ as follows:
\begin{equation}
\Delta t_{up}=\frac{7}{A}\left[ (1-\zeta_0^{2/7})^{1/2}-\frac{2}{3}(1-\zeta_0^{2/7})^{3/2} 
+\frac{1}{5}(1-\zeta_0^{2/7})^{5/2} \right]
\end{equation}
\begin{equation}
\Delta t_{dn}=\frac{7}{A}\left(1 - \sqrt{1-\zeta_0^{2/7}} \right)
\end{equation}
Finally, the pressure over a membrane is described by the piecewise function of the form:
\begin{equation}
p(t)=
\left\{
\begin{array}{lcl}
2p_0(\zeta_{up}(t)-0.5)    & \quad t_0\leq t \leq t_0+\Delta t_{up}\\
p_0,    & \quad t_0+\Delta t_{up} < t < t_0+\Delta t_{up}+\delta t\\
2p_0(\zeta_{dn}(t)-0.5)    &\quad t_0+\Delta t_{up}+\delta t \leq t \leq t_0+\Delta t_{up}+\delta t+\Delta t_{dn}\\
\dots\\
\end{array}
\right.
\label{eq:pressure}
\end{equation}
where $\delta t$ is the time span when the pressure remains constant. We neglect possible deformations of the valve halves after a valve's strap comes into contact with the bottom of the channel and assume that the valve stops completely. Thus, we replace positive values of pressure which bends the membrane with zero, since only the membrane's bending and not its compression makes a contribution to the flow rate. The initial pressure values are assumed to be zero for the convenience of subsequent calculations. At the initial time, all membranes are in the quiescent position. There is no stroke corresponding to this state in the 6-step working algorithm (Fig.~\ref{fig:motion}), so the first ``period'' does not belong to it and is used to enter the normal operating mode. Fig.~\ref{fig:pressures} represents the corresponding pressures over the valve and chamber membrane for $\pm$30 kPa and 3 Hz control unit mode.

\subsection{Membrane motion}
\label{sec:membrane_motion}
The membrane is assumed to be isotropic and shear deformations are neglected. Since the channel height is 100 $\mu$m and the membrane thickness is about 500 $\mu$m, its deflection from the quiescent position is assumed to be small.
An application of a general theory for such circular membrane bending \cite{Reddy2006}, \cite{Singh2002} to the Kelvin-Voigt stress-strain relationship (\ref{eq:KelvinVoigt}) gives the following initial and boundary value problem of the deflection of the membrane midplane from the quiescent position:
\begin{gather}
%\left\{
%\begin{array}{lcl}
 \frac{D}{R^4}\nabla^4 w+\frac{\eta h^3}{12 R^4}\nabla^4 \dot{w}+\rho h \ddot{w}=-p(t)-\gamma\dot{w}
 \label{eq:boundary_problem_eq}\\
 w(1,t)=0  \label{eq:boundary_problem_bc1}\\
 w'(1,t)=0 \label{eq:no_rotation}\\
w'(0,t)=0 \label{eq:finite}\\
 w(\xi,0)=0\label{eq:initial}\\
 \dot{w}(\xi,0)=0
%\end{array}
%\right.
\label{eq:boundary_problem_bc2}
\end{gather}
where $\dot{w}=\partial w/\partial t$, $\nabla^4=\left[\dfrac{1}{\xi}\dfrac{d}{d\xi}\left(\xi\dfrac{d}{d\xi}\right)\right]^2 $, $\xi= r/R$ is the dimensionless radial coordinate, $R$ is the radius of the membrane midplane, $\rho$ is the PDMS density, $h$ is the thickness of the membrane, $\gamma=R_{hyd}\pi R^2$ is the damping coefficient of the channel, $R_{hyd}$ is the hydrodynamic resistance of the channel,
\begin{equation}
D=\frac{Eh^3}{12(1-\nu^2)}
\end{equation}
where $E$ is the PDMS Young modulus, $\nu$ is Poisson's ratio.
%The boundary conditions mean the absence of transverse and rotational displacements on the boundary \cite{Sircar1972}. The membrane displacement is always finite. The initial conditions mean that all membranes were in quiescent position at $t=0$.

In our case, the membrane is made as a step-like reduction in the thickness of the polymer base layer which is rigidly fixed between a glass and polycarbonate, so there is no vertical displacements on its boundary (the condition imposed by Eq. \ref{eq:boundary_problem_bc1}). Since we have no swing joints in our pump, and the membrane material resists fracture and plastic flow, any fracture-like rotational displacements on the boundary could only take place for an infinitely thin membrane (i.e., zero-height plane). The production of such an extremely thin membrane is both impossible and impractical, therefore we always suppose the membrane to have a finite non-zero thickness. But, throughout the solution procedure, we effectively reduce the membrane to its midplane, though still remember that it exhibits the essential properties of a non-zero height object. Thus we have to specify the absence of rotation at the clamping point explicitly by setting the condition (\ref{eq:no_rotation}). The condition imposed by Eq. (\ref{eq:finite}) is needed to keep the displacement of the membrane center finite and avoid non-physical solutions. Finally, the initial conditions (\ref{eq:initial}), (\ref{eq:boundary_problem_bc2}) manifest that all membranes are in quiescent position with zero velocity at $t=0$.

The membrane pushes the liquid through the channel while being deformed. If flow is laminar, the reaction pressure in the channel is proportional to the volume flow rate which depends on the membrane velocity. The damping factor $\gamma$ represents an averaged factor of proportionality between the channel reaction pressure and velocity of the liquid.

The applied pressure $p(t)$ is calculated according to Section~\ref{sec:pressure}.

Equation (\ref{eq:boundary_problem_eq}) with initial and boundary conditions (\ref{eq:boundary_problem_bc1})--(\ref{eq:boundary_problem_bc2}) is solved using the eigenfunction expansion technique. Its solution has the following form:
\begin{equation}
w(\xi,t)=\sum\limits_{n=1}^{+\infty} f_n(\xi)g_n(t)
\label{eq:expansion}
\end{equation}
where
$f_n$ are the eigenfunctions of the corresponding Sturm–Liouville problem:
\begin{gather}
%\begin{cases}
 \nabla^4 f=\lambda^4 f \label{eq:SL_eq}
 \\
 f(1)=0 \label{eq:SL_bc1}\\
 f'(1)=0 \\
 f'(0)=0 \label{eq:SL_bc2}
%\end{cases}
\end{gather}
The solution of this problem has the form $f_n=A_n J_0(\lambda_n\xi)+\tilde{A}_n N_0(\lambda_n\xi) + B_n I_0(\lambda_n\xi)+\tilde{B}_n K_0(\lambda_n\xi)$ where $J_0$ and $N_0$ are Bessel functions, $I_0$ and $K_0$ are modified Bessel functions \cite{Courant1966}. To satisfy the boundary condition $f'(0)=0$, one has to take $\tilde{A}_n=\tilde{B}_n=0$. The remaining boundary conditions are satisfied if
$ B_n=-A_n\dfrac{J_0(\lambda_n)}{I_0(\lambda_n)}$ with eigenvalues $\lambda_n$ being a solution of the following equation:
\begin{equation}
J_0(\lambda_n)I_1(\lambda_n)+I_0(\lambda_n)J_1(\lambda_n)=0
\end{equation}
Hence
\begin{equation}
f_n(\xi)=J_0(\lambda_n\xi)-\frac{J_0(\lambda_n)}{I_0(\lambda_n)}I_0(\lambda_n\xi)
\label{eq:eigenfunc}
\end{equation}

Substitution of expansion (\ref{eq:expansion}) into (\ref{eq:boundary_problem_eq}) yields the following initial value problem to obtain the functions $g_n(t)$:
\begin{gather}
%\left\{
%\begin{array}{lcl}
 {\ddot g_n}+2\beta_n\dot{g_n}+\omega_{0n}^2 g_n=-\frac{L_n}{\rho h}p(t) \label{eq:Cauchy_eq}\\
 g_n(0)=0\label{eq:Cauchy_bc1}\\
 \dot{g_n}(0)=0\label{eq:Cauchy_bc2}
%\end{array}
%\right.
\end{gather}
where
\begin{equation}
\beta_n= \frac{\eta h^2 \lambda_n^4}{24 R^4 \rho} + \frac{\gamma}{2\rho h} \qquad
\omega_{0n}=\sqrt{\frac{D}{\rho h}}\left(\frac{\lambda_n}{R}\right)^2
\end{equation}
\begin{equation}
L_n=\frac{2}{\lambda_n}\frac{J_1(\lambda_n)-\frac{J_0(\lambda_n)}{I_0(\lambda_n)}I_1(\lambda_n) }{J_1^2(\lambda_n) + \left(\frac{J_0(\lambda_n)}{I_0(\lambda_n)}\right)^2 I_1^2(\lambda_n)}
\end{equation}
($L_n p(t)$ are the expansion coefficients of the pressure $p(t)$ in the eigenfunctions $f_n(\xi)$). Equation (\ref{eq:Cauchy_eq}) represents a well-studied damped harmonic oscillator equation with zero initial conditions (\ref{eq:Cauchy_bc1}, \ref{eq:Cauchy_bc2}). Its solution can be found either numerically or using Green's function:
\begin{equation}
g_n(t)=
-\frac{L_n}{\rho h}\int\limits_0^t e^{-\beta_n(t-\tau)}
 \frac{\sin[\omega_n(t-\tau)]}{\omega_n}
 p(\tau)
d\tau
\label{eq:gn}
\end{equation}

Thus, the deflection of the membrane midplane from its quiescent position is described by the following dependency:
\begin{multline}
w(\xi,t)=
-\sum\limits_{n=1}^{+\infty} \frac{L_n}{\rho h}
\left[ J_0(\lambda_n\xi)-\frac{J_0(\lambda_n)}{I_0(\lambda_n)}I_0(\lambda_n\xi) \right]\times\\
\int\limits_0^t e^{-\beta_n(t-\tau)} 
\frac{\sin[\omega_n(t-\tau)]}{\omega_n}
  p(\tau)
 d\tau
\label{eq:displacement}
\end{multline}

\subsection{Volume flow rate}
The volume flow is defined as an integration of the obtained velocity of the membrane over its surface:
\begin{equation}
Q(t)=\int\limits_S \dot{w}(\xi,t)dS
\end{equation}
To solve this integral, curvilinear coordinates are used:
\begin{equation}
x=R\xi\cos\varphi \quad
y=R\xi\sin\varphi \quad
z=w(\xi,t)
\end{equation}
with the following surface area element (\cite{Fichtenholz1965})
\begin{equation}
dS=R^2\xi \sqrt{1+(w'(\xi,t))^2}\,d\xi d\varphi
\end{equation}

Finally, the volume flow rate has the following form:
\begin{equation}
Q(t)=2\pi R\int\limits_0^1
\dot{w}(\xi,t)\xi
\sqrt{R^2+(w'(\xi,t))^2}
d\xi
\label{eq:flowrate}
\end{equation}

The chamber membrane is able to reach the channel bottom while moving downward. The integration lower limit is replaced by $\xi_0(t)$ for this case, being the solution of
\begin{equation}
w(\xi_0,t)=-z_0
\end{equation}
where $z_0$ is the height of the channel.

All the above calculations have been implemented in C++ using the Boost library to work with Bessel functions. We have used the first four expansion terms when calculating $w'$ and $\dot{w}$ since, e.g., the contribution of the third term only affects the fourth decimal point in the value of the average flow rate.

\section{Experimental setup}
\begin{figure}
\begin{center}
\includegraphics{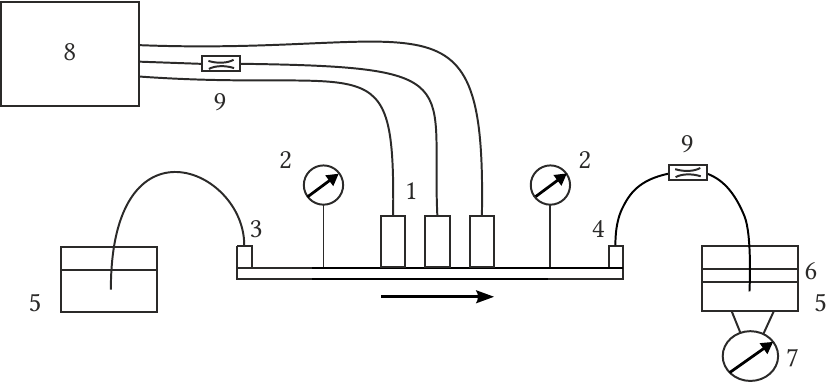}
\caption{The scheme of the experimental setup: 1 -- pump, 2 -- pressure sensors, 3 -- inlet, 4 -- outlet, 5 -- water reservoirs, 6 -- oil layer, 7 -- balance, 8 -- control unit, 9 -- throttles}
\label{fig:experiment}
\end{center}
\end{figure}

The experimental setup is shown schematically in Fig.~\ref{fig:experiment}.
Each pump is connected to the Hemule control unit ``8'' using FESTO PUN-H-2$\times$0.4 pneumatic tubes. The same tubes are used to connect inlet ``3'' and outlet ``4'' of the pump with reservoirs filled with deionized water ``5''. Honeywell 40PC001B pressure sensors ``2'' are placed as shown in Fig.~\ref{fig:pump}. Their readings are captured by a Tektronix MSO 3024 oscilloscope and converted to pressure units according to the sensor datasheet. The obtained pressure difference $\mathrm{\Delta} p(t)$ is converted to the volume flow rate as follows:
\begin{equation}
Q(t)=\mathrm{\Delta} p(t) \frac{Q_{av}}{\frac{1}{2T}\int\limits_{t_0}^{t_0+T}\mathrm{\Delta} p(t)dt}
\end{equation}
where $Q_{av}$ is the average volume flow rate. In order to measure it, the reservoir connected to the outlet of the pump is placed on an OHAUS EX 224 precision balance ``7''. The reservoir is filled with mineral oil ``6'' to prevent the evaporation of the water from the surface. Division by 2 is caused by the fact that the unclosed pump connection scheme is used in the experiment.

The slope of pressure pulses over the membranes is adjusted by FESTO GRLO-M5-QS-3-LF-C throttles ``9'' mounted on a pneumatic tubes. The same throttle is used on the outlet tube to simulate the adjustment of friction loss of the microfluidic channel.

\section{Results and discussion}
\begin{figure}[!ht]
\begin{center}
\includegraphics[width=\textwidth]{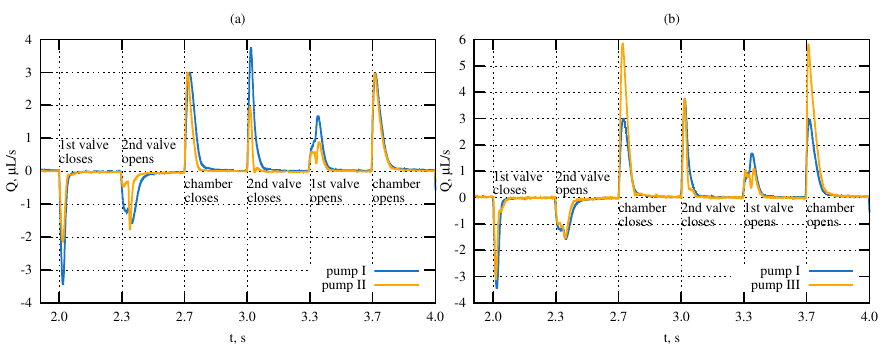}
\caption{Comparison of volume flow rates generated by (a) pump I and pump II (smaller valves) and (b) pump I and pump III (thinner working chamber membrane)}
\label{fig:experiment_comparison}
\end{center}
\end{figure}

\begin{figure}[!ht]
\begin{center}
\includegraphics[width=\textwidth]{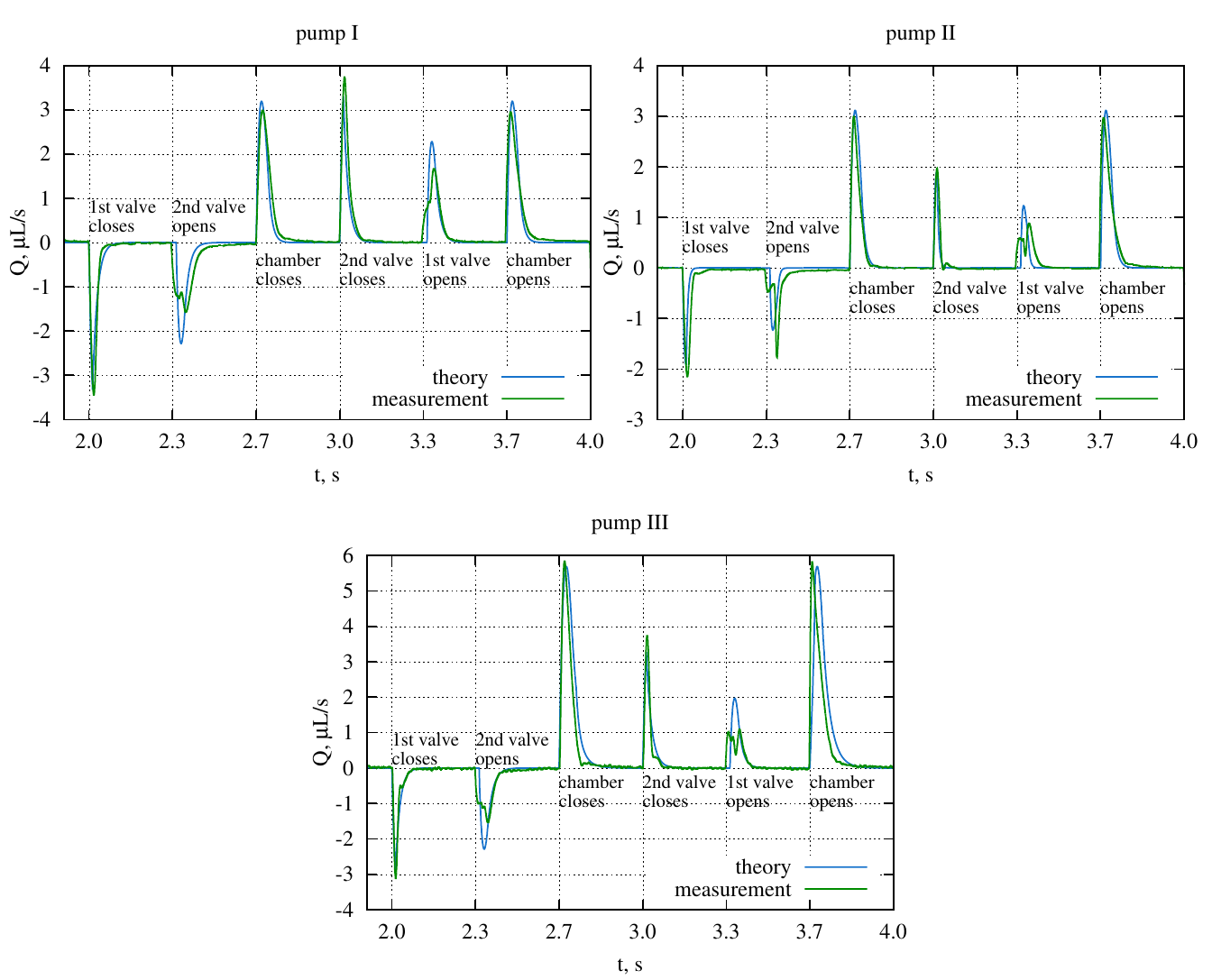}
\caption{Comparison between the experimental results and the theoretical predictions}
\label{fig:theory_experiment}
\end{center}
\end{figure}

Measurements of the volume flow rate have been carried out for all versions of the pump. The hydrodynamic resistances are estimated according to \cite{Bruus2008} and are about 280~Pa$\cdot$s/mm$^3$ for a channel part between an outlet and a chamber and about 220~Pa$\cdot$s/mm$^3$ for a part between an outlet and a valve. PDMS viscosity is taken to be 32~kPa$\cdot$s \cite{Swallow2002}. 
Experimental results (Fig.~\ref{fig:experiment_comparison}) show that valves with a smaller diameter produce peaks with a smaller magnitude (pump I and pump II), whereas the working chamber with a thinner membrane produces a higher peak (pump I and pump III). Negative values of the volume flow rate indicate a backflow. If the pump resides inside a closed microchannel, both opening the first valve and closing the second valve causes the fluid to move in the opposite direction. Theoretical predictions are in close agreement with experimental results (Fig.~\ref{fig:theory_experiment}). Some discrepancy between the theory and the experiment is observed when a valve opens. It happens probably due to the non-laminar transitional flow when the strap of a valve is detached from the bottom of the channel and the fluid flows under it. Consideration of this transition process is beyond the scope of this paper.

\begin{figure}[!ht]
\begin{center}
\includegraphics[width=\textwidth]{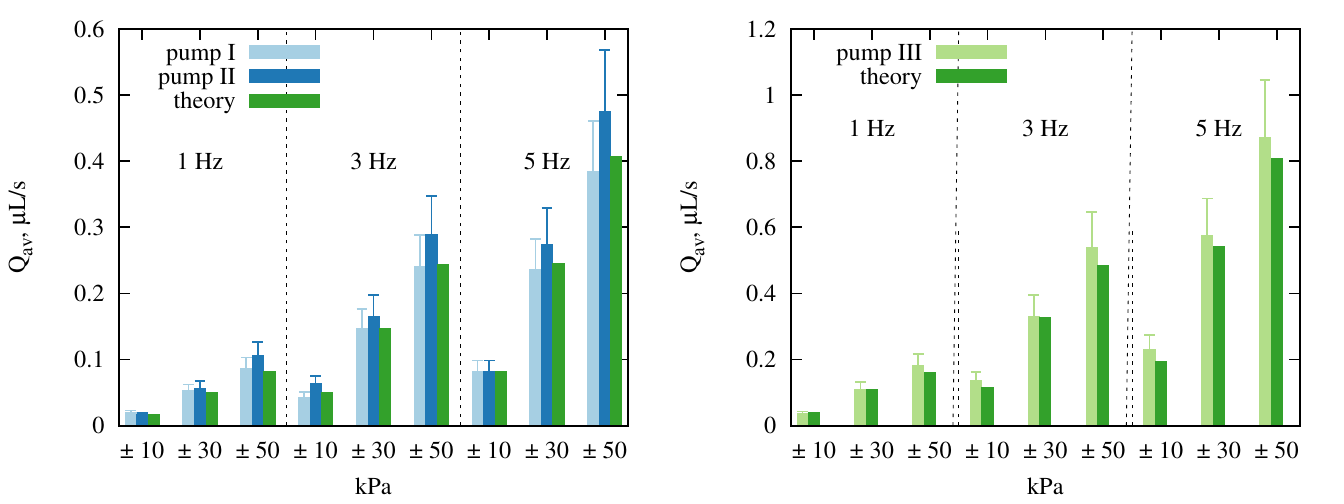}
\caption{The average volume flow rate of investigated pumps compared with theory}
\label{fig:average_flow_rate}
\end{center}
\end{figure}

Average volume flow rates over one pumping cycle for all types of the pump and for different control unit modes are represented in Fig.~\ref{fig:average_flow_rate}. As follows from these results, the average volume flow rate is proportional to the pumping frequency and the pressure magnitude until the chamber membrane reaches the channel bottom. The theoretically predicted average volume flow rates are equal for pump I and pump II because the contributions of valves compensate for each other if the valve properties are identical. The measured data coincides within a measurement error.

\begin{figure}
\includegraphics[width=\textwidth]{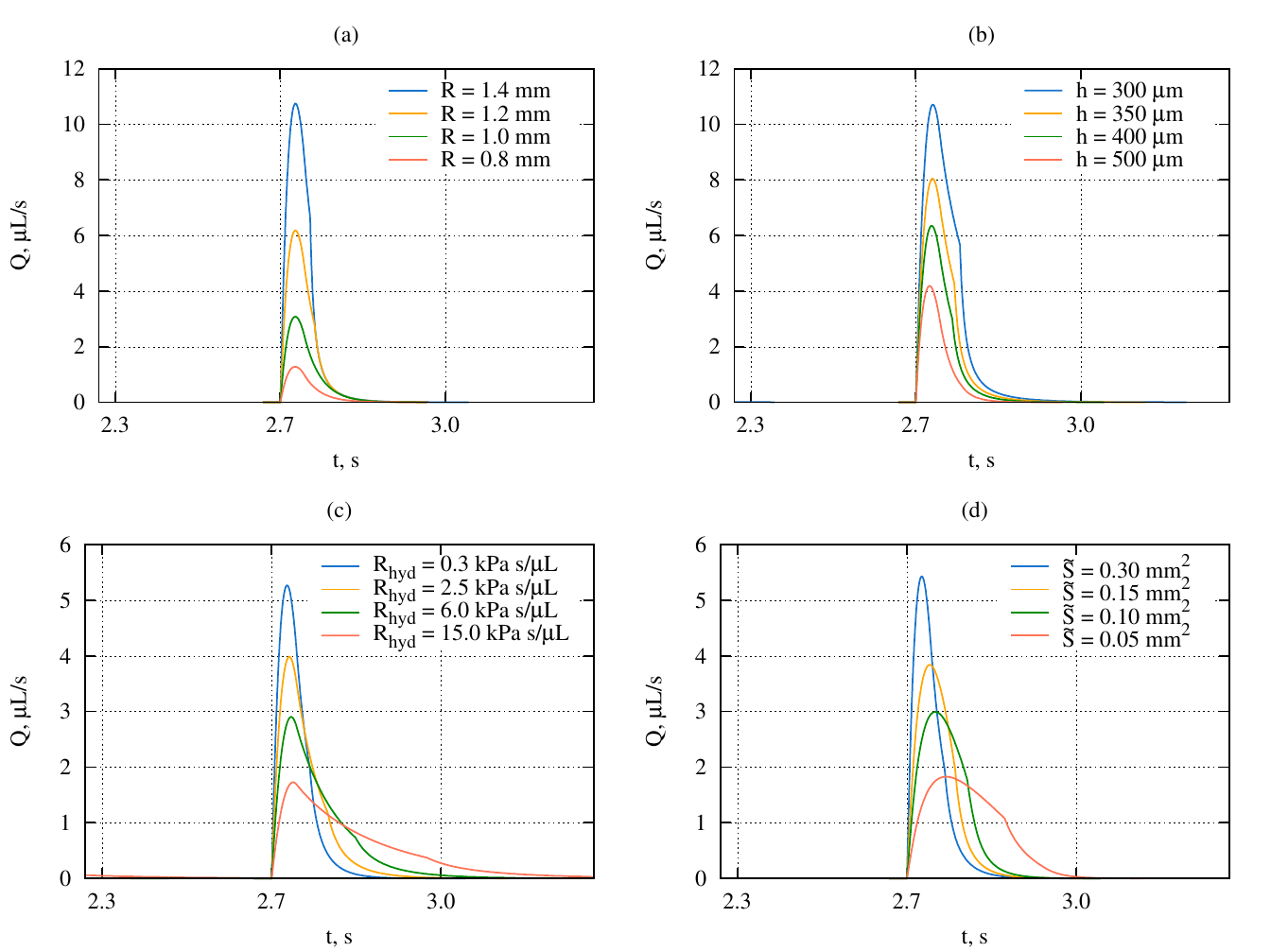}
\caption{The impact of a membrane radius (a), its thickness (b), hydrodynamic resistance of a channel (c) and the slope of the pressure pulse (determined by $\tilde{S}$) on the form of the peak created by the chamber of pump III} 
\label{fig:theory_parameters} 
\end{figure}

Fig.~\ref{fig:theory_parameters} represents theoretical dependencies of flow rate peaks created by the chamber membrane with different pump parameters. The investigated parameters are the membrane radius (Fig.~\ref{fig:theory_parameters}a) and thickness (Fig.~\ref{fig:theory_parameters}b), as well as the hydraulic resistance of the channel after the pump (Fig.~\ref{fig:theory_parameters}c) and the slope of the air pressure pulse over the membrane (Fig.~\ref{fig:theory_parameters}d). The last is governed by throttles connected to the control unit outputs. (The smaller the throttle orifice, the gentler the slope.)

\begin{figure}[ht!]
\begin{center}
\includegraphics{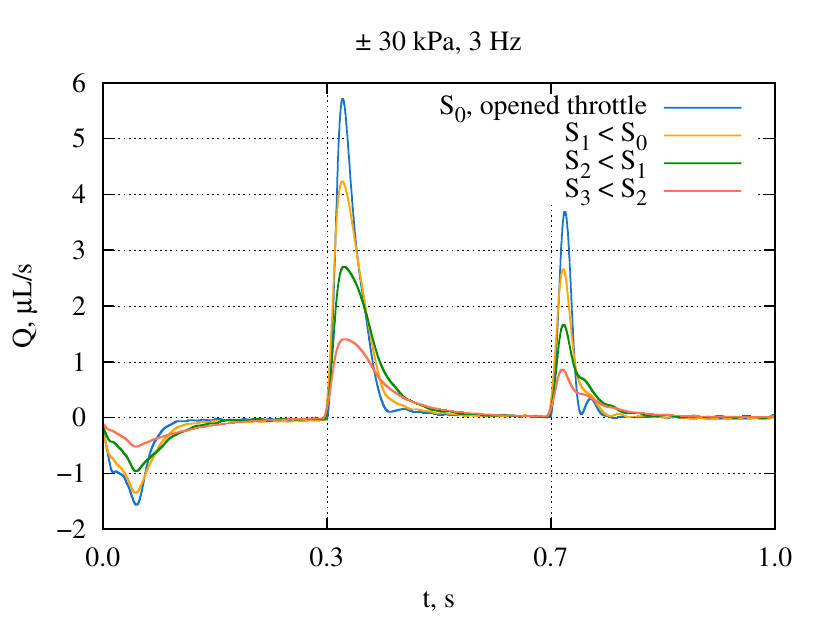}
\caption{The throttle is mounted on the outlet to increase the hydrodynamic resistance after the pump. The Figure shows peaks obtained from the outlet pressure sensor for different throttle cross-sections.}
\label{fig:water_throttle}
\end{center}
\end{figure}

A curve break corresponds to the membrane reaching the bottom of the channel. 
The experimental evidence for the influence of the membrane radius and thickness on the peak form can be seen in Figs.~\ref{fig:experiment_comparison} and \ref{fig:theory_experiment}. The qualitative experimental dependency of the peak on the hydrodynamic resistance of a channel is shown in Fig.~\ref{fig:water_throttle}. The hydrodynamic resistance is changed by varying the cross-section of the throttle connected to the outlet. Such throttle does not influence data obtained from a sensor placed closer to the inlet, therefore only the data from the sensor placed closer to the outlet is shown. The obtained curve represents the contribution to the flow rate of a valve nearest to the sensor and a chamber when this valve is open.  

\begin{figure}[ht!]
\begin{center}
\includegraphics{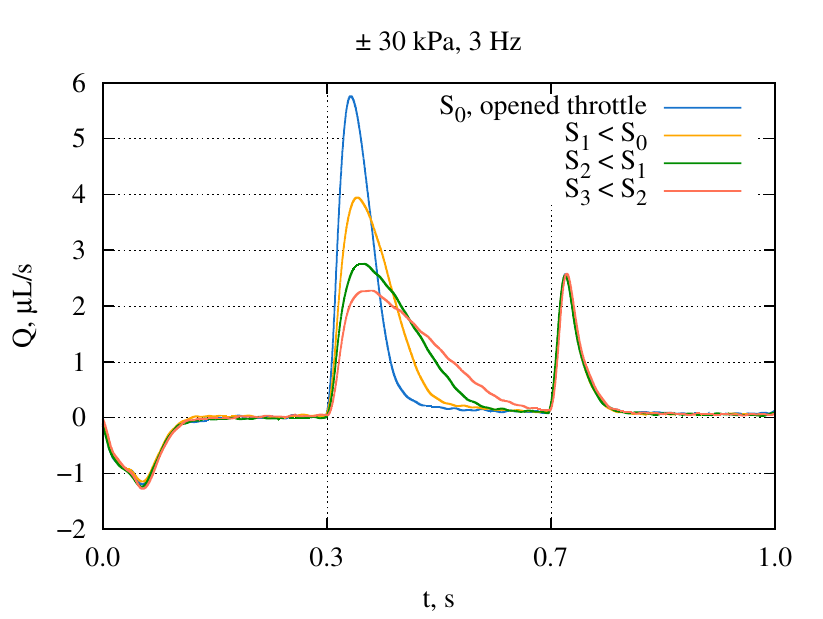}
\caption{
The slope of the pressure pulse is changed by varying the control unit throttle cross-section. The Figure shows peaks obtained from the outlet pressure sensor for different throttle cross-sections.}
\label{fig:air_throttle}
\end{center}
\end{figure}

Similar measurements are carried out in order to study the dependency of the peak on the pressure slope. The latter is determined by a pneumatic tube orifice (Section~\ref{sec:pressure}) and is changed with a throttle mounted on a pneumatic tube that operates the pressure over the membrane of the chamber. The corresponding data from a sensor placed closer to the pump outlet is shown in Fig.~\ref{fig:air_throttle}.
All measurements have been made using pump III.

In the general case, an exact solution of the fluid flow problem in a microchannel under the action of elastic membranes requires jointly solving the Navier-Stokes and the solid-state deformation equations (FSI), so this is only possible numerically. In this work, we separated the fluid flow problem from the problem of membrane deformation by introducing a damping coefficient which takes into account the hydraulic resistance of the channel in a general form.

The proposed model differs from the previously known models of a similar pump \cite{Nedelcu2007, Busek2013} by a more accurate description of the pressure variation over the membranes (allowance for the presence of a throttle, adiabaticity of the gas flow process), as well as by the use of the equations of dynamics for describing the membranes' motion. The simulation results are limited to the linear deformation region of PDMS (strain up to 45\% \cite{Schneider2009} and the maximum displacement $w_0$ less than the membrane's thickness $h$ \cite{Timoshenko1959}). Taking into account the viscoelastic properties of PDMS allows us to obtain the correct width of the volume flow rate peaks (Figs.~\ref{fig:experiment_comparison} and \ref{fig:theory_experiment}). If PDMS is considered to be a Hookean material, the width of the peaks should be equal to the pressure rise/fall time, which, according to the experiment, is tens of milliseconds. At the same time, the width actually observed in the experiment turns out to be much larger (hundreds of milliseconds).

Besides that, a number of assumptions have been made in order to simplify the solution procedure. Namely, the channel walls are assumed to be stiff, and the valves are expected to stop completely when their straps reached the bottom of the channel. The problem of contact between the membrane of the working chamber and the chamber bottom cannot be solved analytically, so when calculating the volume flow rate, we assign a zero velocity to all membrane points whose calculated displacement exceeded the height of the channel. This simplification make it possible to preserve the analyticity of the solution. Comparison with experiment shows that this approximation does not introduce a significant error.

Mapping the membrane traction to the flow rate, the technique we used, is not suitable in all cases when the algorithm does not involve a complete blocking of the channel on one side at every time, and the fluid can flow in both directions. In this more general case, one has to calculate the flow rate using the fluid velocity profile, assuming that the law of motion of the channel boundary $w(\xi,t)$ is known. Numerical solution of such problem either with the help of computational hydrodynamics methods or by means of meshfree computations proposed in \cite{Aboelkassem2012, Aboelkassem2015} is the goal of further research.

\section{Conclusion}
The mathematical model presented in this paper provides the volume flow rate generated by a peristaltic-like displacement micropump with active valves by a straightforward description of the pump membranes' motion. The model has been successfully validated by experiment with three versions of the micropump made in a thin polydimethylsiloxane (PDMS) layer.
In our setup, the maximum average output flow rate of 0.9 $\mu$L/s is achieved by the optimal design PDMS membrane of  2.5 mm in diameter and 440 $\mu$m in thickness at the operating frequency of 5 Hz and pressure of 50 kPa. Small valve membranes of 2 mm in diameter are used to reduce backflow.

The main advantage of the developed model is that, in contrast to models already known in the literature, it only uses the physical parameters of the pump and allows the estimation of their influence on the resulting flow, which is crucially important at the pump design stage. At the same time, it does not require sophisticated software for implementation.
 
The model can be used to describe a variety of pumps working on a similar principle. Since the calculation performed in Section~\ref{sec:membrane_motion} do not use an explicit form of the applied pressure $p(t)$, one can substitute any specific force into the right-hand side of the equation and thus use this model to describe pumps with different actuation types.

Since, in general, our model is aimed to theoretically predict the characteristics of peristaltic-like displacement micropumps and thus provide assistance in micropump device development process, we believe that our results can be useful for both engineers and researchers working in the area of microfluidics.

\section{Acknowledgments}
We are immensely grateful to Dr. Timur R. Samatov from Evotec International GmbH, Dr. Igor E. Frolov from M.V. Lomonosov Moscow State University and Dr. Alexander V. Tyukov from University of Southern California for their invaluable help during article preparation. 

\bibliography{gerasimenko_kindeeva}

\end{document}